\documentclass[%
 reprint,
 amsmath,amssymb,
 aps
]{revtex4-2}

\usepackage{graphicx}
\usepackage{dcolumn}
\usepackage{bm}
\usepackage[utf8]{inputenc}
\usepackage[T1]{fontenc}
\usepackage{amsmath}
\usepackage{amsfonts}
\usepackage{amssymb}
\usepackage[colorlinks=true,citecolor=blue,urlcolor=magenta]{hyperref}
\usepackage{adjustbox}
\usepackage{mathtools}
\usepackage{xfrac}
\hypersetup{breaklinks=true} 

\usepackage[english]{babel}
\usepackage[autostyle, english = american]{csquotes}
\MakeOuterQuote{"}

\newcommand{\ket}[1]{\left| #1 \right>}

\newcommand{\ketbra}[2]{\left| #1 \right>\left< #2 \right|}
\DeclareMathOperator{\Tr}{Tr}
\DeclarePairedDelimiter\ceil{\lceil}{\rceil}

\newcommand{\Atensor}{A(j_1,\ldots,j_Q)}
\newcommand{\Gate}[1]{G^{[#1, d_{#1}]}}
\newcommand{\GateMatrix}[3]{G^{[#1, d_{#1}]}\left(#2;#3\right)}

\begin{document}

\preprint{APS/123-QED}

\title{Deterministic and entanglement-efficient preparation of amplitude-encoded quantum registers}

\author{Prithvi Gundlapalli}

\author{Junyi Lee}
 \email{lee_jun_yi@imre.a-star.edu.sg}
\affiliation{Institute of Materials Research and Engineering, Agency for Science, Technology and Research (A*STAR), 2 Fusionopolis Way, \#08-03 Innovis, 138634 Singapore}

\date{\today}

\begin{abstract}
Quantum computing promises to provide exponential speed-ups to certain classes of problems. In many such algorithms, a classical vector $\mathbf{b}$ is encoded in the amplitudes of a quantum state $\ket{b}$. However, efficiently preparing $\ket{b}$ is known to be a difficult problem because an arbitrary state of $Q$ qubits generally requires $\sim 2^Q$ entangling gates, which results in significant decoherence on today’s Noisy-Intermediate Scale Quantum (NISQ) computers. We present a deterministic (non-variational) algorithm that allows one to flexibly reduce the quantum resources required for state preparation in an entanglement-efficient manner. Although this comes at the expense of reduced theoretical fidelity, actual fidelities on current NISQ computers might actually be higher due to reduced decoherence. We show this to be true for various cases of interest such as the normal and log-normal distributions. For low entanglement states, our algorithm can prepare states with more than an order of magnitude fewer entangling gates as compared to isometric decomposition.
\end{abstract}

\maketitle


\section{\label{sec:intro}Introduction}

The inefficient preparation of quantum states is a bottleneck that currently prevents many quantum algorithms from achieving quantum supremacy over their classical counterparts. For example, the exponential speed-up of the Harrow-Hassidim-Lloyd (HHL) \cite{Harrow_2009} algorithm presumes the existence of an efficient way to encode the components of a classical vector $\mathbf{b}$ in the amplitudes of a quantum state $\ket{b}$. However, such an oracle remains elusive. Beyond HHL, amplitude encoding is used in many other diverse areas including singular value estimation \cite{Wossnig_2018}, least-squares fitting \cite{Wiebe_2012}, and machine learning such as support vector machines \cite{Rebentrost_2014} and autoencoders \cite{Romero_2017}. All these algorithms presume accurate state preparation but circuits that are overly deep threaten to introduce unacceptable decoherence in today's NISQ computers \cite{Preskill_2018}, resulting in an unusable set of states that hinders correct execution of the main algorithm. In this work, we demonstrate a state preparation algorithm that can allow states to be prepared with higher fidelities on current NISQ computers by allowing the user to, if necessary, optimally prepare the state with fewer quantum resources, which can result in improved fidelities on actual NISQ computers due to reduced experimental decoherence.

Over the years, numerous state preparation schemes have been proposed, including preparation via factorization of the Hilbert space \cite{Plesch_2011}, decomposition of isometries \cite{Iten_2016}, quantum generative adversarial networks (qGAN) \cite{Zoufal_2019, Situ_2020}, variational quantum circuits \cite{Ho_2019, Cincio_2021}, and matrix product states (MPS) inspired approaches \cite{Ran_2020, holmes2020efficient}. However, these methods all suffer from various problems and are less than ideal. Variational methods, like qGAN and most variational quantum circuits, use a fixed circuit ansatz with parametrized gates that are progressively tuned through iterative optimization steps to prepare the target state. Although these approaches have enjoyed some success in preparing relatively small quantum registers ($\sim$ 3 qubits), their general applicability is still doubtful for several reasons. Firstly, optimization of these circuits requires the use of a loss function, but it is known that the gradient of many loss functions vanish exponentially for large system sizes (the "barren plateau" problem), making it extremely difficult for the optimizer to make any progress \cite{Cerezo_2021}. Secondly, besides vanishing gradients, numerical experiments also suggest that the loss function's landscape is pockmarked with a multitude of local minima, making the optimizer's task even more challenging \cite{Day_2019}. Thirdly, there is no compelling reason to choose one circuit topology over another and since there is no guarantee that the chosen ansatz will eventually yield the desired state, it is difficult to know when to stop optimizing and to try a different topology instead. While attempts have been made to also optimize the circuit's topology \cite{Cincio_2018, Cincio_2021}, it is likely that this will only aggravate the severeness of the optimization problem when scaled to larger system sizes. Finally, we note that in all these variational approaches, there is an unspecified cost of classical training time, which could well negate the theorized speed-up of quantum algorithms.

In contrast to these variational methods, there are other analytical approaches that do not require intense optimizations and whose results are mathematically guaranteed (in an ideal noiseless quantum computer). They are, in that sense, "deterministic" as compared to variational methods where there is no guarantee of a solution's existence given a specified circuit topology. The chief disadvantage of these methods is that exact preparation of an arbitrary $Q$-qubit state generally requires $\sim 2^Q$ entangling gates \cite{Bergholm_2005, Plesch_2011, Iten_2016}, which is prohibitively expensive in the NISQ era. More recently, there have been proposals to shorten the depth of these circuits by utilizing additional ancillary qubits \cite{Araujo_2021}. Nevertheless, it should be noted that the final state of such routines is entangled with the ancillary qubits. Consequently, it cannot be accepted by all algorithms expecting a usual amplitude encoding without additional modifications or measurements.

A common deficiency of all the state preparation routines discussed thus far is their insensitivity to the entanglement of the target state. By this we mean that although a generic $Q$-qubit state may require an exponential number of entangling gates to prepare, it is clear that not every state requires such extensive resources to prepare. Indeed, a product state with no entanglement can be prepared with a single layer of non-entangling single qubit gates, but current deterministic approaches do not explicitly account for this difference between states and can, as we show in sections \ref{sec:hardware_benchmark} and \ref{sec:simulation_benchmark}, potentially use an order of magnitude more entangling gates than necessary. Furthermore, although a highly entangled state may require a substantial number of entangling gates to prepare, it is worth asking if one might find an optimally approximate state (in the amplitudes of the state) that has lower entanglement, and is therefore easier to prepare with fewer entangling gates.

In this paper, we utilize the properties of matrix product states to prepare arbitrary states with varying resources depending on the desired fidelity. Although our algorithm can prepare a target state exactly, its main utility in the NISQ era is to prepare an approximate state with fewer resources because one might ironically achieve higher fidelity to the target state by preparing an approximate state than by preparing an exact state that requires significantly more entangling gates (see, for example, our tests on actual NISQ computers in section~\ref{sec:hardware_benchmark} and the results in Table~\ref{tab:benchmark}). Compared to other matrix product state based circuits that use $Q - 1$ sequential fixed $d$ qubit gates \cite{Cramer2010}, or multiple layers of fixed 2-qubit gates \cite{Ran_2020}, our approach uses one layer of $Q - 1$ sequential $d_q$ qubit gates, where $d_q$ is a variable that depends on the entanglement of the target state and $q = 1,\ldots,Q - 1$ for a $Q$ qubit system. In the case of a separable state, our algorithm automatically reduces to a single layer of single qubit gates i.e. $d_q = 1$ for all $q$. As we show in section~\ref{sec:simulation_benchmark}, this flexibility of using variable sized gates can allow for more efficient state preparation compared to having multiple layers of 2-qubit gates, and can frequently prepare states with more than an order of magnitude fewer gates compared to isometric decomposition, which is to our knowledge one of the most efficient deterministic state preparation algorithm available requiring to leading order $2^Q$ CNOT gates for preparing an arbitrary $Q$ qubit state. This performance is comparable to $\sfrac{23}{24}\,\, 2^Q$ CNOT gates for even $Q$ \cite{Plesch_2011}, $2^Q$ CNOT gates \cite{Bergholm_2005}, and better than $2^{Q + 1}$ CNOT gates \cite{Shende2006}. During preparation of this manuscript, we were made aware of a similar circuit utilizing variable sized gates in the appendix of a recently published work, although the authors there did not study the use of such a circuit for state preparation \cite{Lin2021}.

The rest of the paper is structured as follows: in section~\ref{sec:matrix_product_states}, we review relevant properties of matrix product states, and in section \ref{sec:matrix_product_initializer}, we describe the details of our state preparation algorithm. We then demonstrate how our algorithm can outperform isometric state preparation on current NISQ computers in section~\ref{sec:hardware_benchmark} and systematically compare our method against other approaches in numerical simulations of 8, 12, and 16 qubit registers in section~\ref{sec:simulation_benchmark}. Finally, we conclude in section~\ref{sec:conclusion}.


\section{\label{sec:matrix_product_states}Matrix Product States}

Matrix product states, or tensor-trains as they are known in computational mathematics literature \cite{Holtz_2012, Oseledets_2011}, are used widely in many areas including generative modeling in machine learning \cite{Han_2018}, estimation of singular values \cite{Lee_2015}, studying of phase transitions in field theories \cite{Milsted_2013}, and perhaps most famously, in the density matrix renormalization group technique for studying 1-dimensional quantum many-body systems \cite{Schollwock_2005}. At the heart of a MPS is the re-writing of the expansion coefficients $A(j_1,\ldots,j_Q)$ of the $Q$-body state $\sum_{j_1,\ldots,j_Q} A(j_1,\ldots,j_Q) \ket{j_Q,\ldots,j_1}$ into a sum-product of $Q$, 3-dimensional tensors $A^n_{\alpha^{n-1},j_n,\alpha^n}$ (also sometimes referred to as a MPS \textit{core})
\begin{align}
    \nonumber
    \ket{\psi} = \sum_{\substack{j_1\ldots j_Q \\ \alpha^0\ldots\alpha^Q}} A^1_{\alpha^0,j_1,\alpha^1} A^2_{\alpha^1,j_2,\alpha^2} \ldots &A^Q_{\alpha^{Q-1},j_Q,\alpha^Q} \\
    &\times \ket{j_Q, \ldots, j_1},
\label{eq:mps_def}
\end{align}
%
where the $\alpha^n$ are summation indices (also known as \textit{virtual} or \textit{bond} indices) with dimensions $\dim(\alpha^n) \in \mathbb{N}$ for $1 \leq n < Q$, $n \in \mathbb{N}$, and $\dim(\alpha^0)=\dim(\alpha^Q)=1$. $j_n$ here refers to the computational basis state of the $n$\textsuperscript{th} qubit, and $\dim(j_n)$ is therefore always 2. More generally, we shall use lowercase latin alphabets to represent physical indices and greek letters to label bond indices. Numerical superscripts are used to enumerate the $Q + 1$ different bond indices while numerical subscripts shall identify a particular qubit. We shall also refer to the $\alpha^{n-1}$ and $\alpha^{n}$ bond indices of $A^n_{\alpha^{n-1},j_n,\alpha^n}$ as the \textit{left} and \textit{right} bond indices of $A^n$ respectively.


Analytical decomposition of the expansion coefficients $\Atensor$ into a sum-product of MPS cores are known for only a few special cases \cite{Oseledets_2013}, but decomposition of $A$ can always be accomplished numerically via successive singular-value decompositions (SVD) \cite{Vidal_2003, Oseledets_2011}. Since we wish to emphasize that our approach works for arbitrary quantum states, we briefly illustrate how this is accomplished. Moreover, it highlights a simple way in which locally optimal approximations to the target state can be systematically achieved.

We begin by noting that the expansion coefficients $\Atensor$ can always be reshaped into a matrix. We shall use the notation $\Atensor$ to denote $A$ as a $Q$-dimensional tensor, and $A(j_1,\ldots,j_n;j_{n+1},\ldots,j_Q)$ as a matrix with $\prod_{i=1}^n \dim(j_i)$ rows and $\prod_{i=n+1}^Q \dim(j_i)$ columns. Given the expansion coefficients $\Atensor$, we then perform the following series of SVD and reshaping operations
\begin{align}
    \nonumber
    &\Atensor = \sum_{\alpha^1} U(j_1;\alpha^1) B(\alpha^1;j_2,j_3,\ldots,j_Q) \\
    \nonumber
    &\quad= \sum_{\alpha^1} A^1_{\alpha^0,j_1,\alpha^1} B(\alpha^1,j_2;j_3,\ldots,j_Q) \\
    \nonumber
    &\quad= \sum_{\alpha^1\alpha^2} A^1_{\alpha^0,j_1,\alpha^1} U(\alpha^1, j_2;\alpha^2) B(\alpha^2;j_3,\ldots,j_Q) \\
    \nonumber
    &\quad= \sum_{\alpha^1\alpha^2} A^1_{\alpha^0,j_1,\alpha^1} A^2_{\alpha^1, j_2,\alpha_2} B(\alpha^2,j_3;j_4,\ldots,j_Q) \\
    \nonumber
    &\quad\quad\quad\quad\quad\quad\quad\quad\quad \vdots \\
    &\quad= \sum_{\alpha^0\ldots\alpha^Q} A^1_{\alpha^0,j_1,\alpha^1} A^2_{\alpha^1,j_2,\alpha^2} \ldots A^Q_{\alpha^{Q-1},j_Q,\alpha^Q}.
\label{eq:TT_SVD}
\end{align}
In the first line, we have reshaped $\Atensor$ into a matrix $B(j_1;j_2,\ldots,j_Q)$ and performed a SVD so that $B(j_1;j_2,\ldots,j_Q) = \sum_{\alpha^1} U(j_1;\alpha^1) B(\alpha^1;j_2,j_3,\ldots,j_Q)$. Notice that $\dim(\alpha^1)$ is the number of columns of $U$, and is also the number of singular values. Next, we reshape $U(j_1;\alpha^1)$ into the first of our desired MPS core $A_{\alpha^0,j_1,\alpha^1}$ (recall that $\dim(\alpha^0) = 1$), and the matrix $B(\alpha^1;j_2,j_3,\ldots,j_Q)$ with $\dim(\alpha^1)$ rows into a matrix $B(\alpha^1,j_2;j_3,\ldots,j_Q)$ with $\dim(\alpha^1)\times\dim(j_2)$ rows. We then perform a SVD on $B(\alpha^1,j_2;j_3,\ldots,j_Q)$ in the third line, and an analogous reshaping operation on the fourth line. This process is then iterated until $\Atensor$ is fully decomposed into $Q$ MPS cores as desired.

This explicit decomposition suggests a natural and easy way to make locally optimal approximations: at each $n$\textsuperscript{th} SVD step, keep only the $k_n$ largest singular values out of all $\dim(\alpha^n)=r_n$ singular values
\begin{align}
    \nonumber
    &B(\alpha^{n-1},j_n;j_{n+1},\ldots,j_Q) \equiv B_n \\
    \nonumber
    &= \sum_{\alpha^n=1}^{r_n} U(\alpha^{n-1}j_n;\alpha^n) B(\alpha^n;j_{n+1},\ldots,j_Q) \\
    &\approx \sum_{\alpha^n=1}^{k_n} U(\alpha^{n-1}j_n;\alpha^n) B(\alpha^n;j_{n+1},\ldots,j_Q) \equiv B_{k_n}.
\label{eq:svd_approx}
\end{align}
%
%
From the Eckart-Young theorem \cite{Eckart_1936}, this is locally optimal in the sense that $B_{k_n}$ is the best $k_n$ rank approximation to $B_n$ (as defined in Eq. \eqref{eq:svd_approx}) under both the Frobenius and spectral norm, with the error in the Frobenius norm simply given by the quadrature sum of the truncated singular values. 
Physically, we should understand these truncations as effectively finding optimally approximate states (in the Frobenius norm of the expansion coeffcients $\Atensor$) that have lower entanglement than the original state. 


Lastly, we note that the decomposition of $\Atensor$ into the MPS form is not unique. One easy way to see this is to recognize that at each SVD in Eq. \eqref{eq:TT_SVD}, we may insert the identity $I=X X^{-1}$ (where $X$ is any unitary matrix) in between $U$ and $B$ and re-define $U X \to U$, $X^{-1} B \to B$ such that each $A^n_{\alpha^{n-1},j_n,\alpha^n}$ in the decomposition can be changed without altering the quantum state. This gauge freedom allows a MPS state to be transformed such that the cores $A^n$ for $1 < n \leq Q$ obey the orthonormal relations
\begin{equation}
    \sum_{j_n, \alpha^n} A^n_{\alpha^{n-1},j_n,\alpha^n} A^{n\,*}_{\beta^{n-1},j_n,\alpha^n} = \delta_{\alpha^{n-1},\beta^{n-1}}.
    \label{eq:right_norm_def1}
\end{equation}
Moreover, if the MPS state is normalized, $A^1$ also obeys the relation
\begin{equation}
    \sum_{j_1,\alpha^1} A^1_{\alpha^{0},j_1,\alpha^1} A^{1\,*}_{\alpha^{0},j_1,\alpha^1} = 1.
    \label{eq:right_norm_def2}
\end{equation}
A MPS state that obeys these relations is said to be a \textit{right-canonical} MPS \cite{Paeckel_2019} and we always shall assume that our MPS state is in the right-canonical form. Our requirement that the MPS state is right-normalizable places constraints on the order of SVD truncations when approximating a state. In particular, the dimension of the left bond index of $A^n$ in any right-normalizable MPS must obey the inequality
\begin{equation}
    \dim(\alpha^{n-1}) \leq \dim(j_n)\dim(\alpha^n) = 2 \dim(\alpha^n).
    \label{eq:dim_inequality}
\end{equation}
To see why this is true, we note that the right normalization relations (Eq. \eqref{eq:right_norm_def1} and \eqref{eq:right_norm_def2}) can be interpreted to mean that each $A^n$ core has $\dim(\alpha^{n-1})$ orthogonal $\mathbb{C}^{\dim(j_n)\times\dim(\alpha^n)}$ vectors. Since a $\mathbb{C}^{\dim(j_n)\times\dim(\alpha^n)}$ vector space cannot have more than $\dim(j_n)\times\dim(\alpha^n)$ orthogonal vectors, the inequality Eq. \eqref{eq:dim_inequality} naturally follows. 

Given a $Q$ qubit system, there are $Q - 1$ non-trivial SVD to perform. To obtain the next best approximation, the smallest singular value, relative to the largest singular value in the same SVD, should be dropped. However, inequality Eq. \eqref{eq:dim_inequality} implies that this should only be done when the resulting MPS obeys Eq. \eqref{eq:dim_inequality}. We note that this only affects the \textit{order} of the truncations; ultimately, any MPS state can be truncated all the way down such that $\dim(\alpha^n) = 1$ for all $n$.

\section{\label{sec:matrix_product_initializer}Matrix Product Initializer}

We now show how to prepare an arbitrary quantum state that has been expressed as a right-canonical MPS. Previously, Ran showed how a MPS consisting of cores with bond dimensions of 2 may be prepared exactly with sequential 2-qubit gates \cite{Ran_2020}. However, this sequence was unable to exactly prepare MPSs with cores of arbitrary bond dimensions, which limits the usefulness of the technique. Here, we show how MPSs with cores of arbitrary bond dimensions may be prepared, which implies that any arbitrary state may be prepared using our technique. More importantly, it also allows for optimally approximate states with lower entanglement than the original state to be prepared, which can be very useful in the current NISQ era.

\begin{figure}[!ht]
\includegraphics[width=0.49\textwidth]{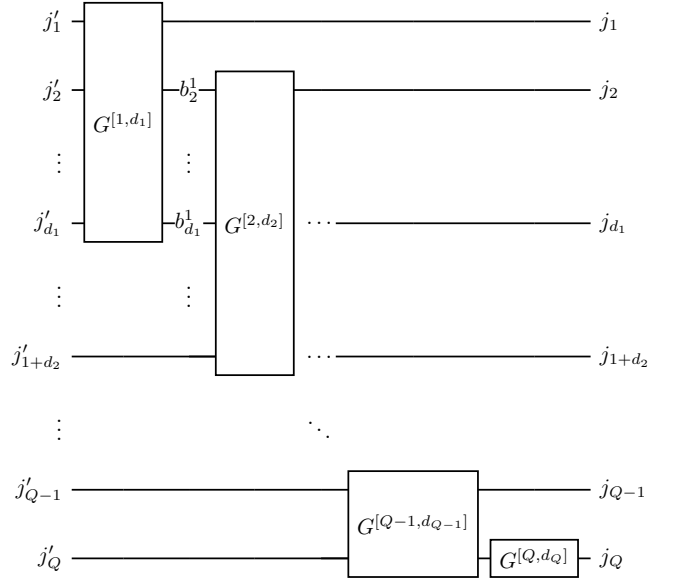}
\caption{Matrix product initializer ansatz for $Q$ qubits consisting of $Q$ sequential (potentially multi-qubit) gates. Each gate $G^{[n, d_n]}$ is a $d_n$-qubit gate operating on the $n$\textsuperscript{th} to $n + d_n - 1$\textsuperscript{th} qubit, where $d_n$ depends on the state's entanglement. $j_n'$ are input states, $j_n$ are the final output states, and $b^m_n$ are intermediate states.}

\label{fig:circuit_ansatz}
\end{figure}

The problem of state preparation may be described as finding a unitary matrix $U$ such that $U\ket{0,\ldots,0}=\ket{\psi}$, where $\ket{\psi}$ is the desired target state to prepare. There is however no unique solution to $U$. On a gate-based quantum computer, $U$ must be formed from the product of several $d$-qubit gates, where $d \in \mathbb{N}, d \leq Q$. However, multi-qubit gates are experimentally difficult to implement and error-prone in the NISQ era. We therefore seek to find a $U$ that only uses $d > 1$ qubit gates when "absolutely necessary". By "absolutely necessary", we mean that in general, $d > 1$ qubit gates are only needed when there is significant entanglement in the target quantum state. After all, a separable product state can be prepared with just one layer of 1 qubit gates. We thus want a $U$ that is made out of variable $d$ qubit gates with $d > 1$ only when the entanglement of the state calls for it. Moreover, it is plain that every qubit needs at least one gate to operate on it. Consequently, we write down as an ansatz that $U$ can be given by the quantum circuit that is shown in Fig. \ref{fig:circuit_ansatz}. Essentially, the ansatz consists of $Q$ sequential $d_n$-qubit gates operating on the $n$\textsuperscript{th} to $n + d_n - 1$\textsuperscript{th} qubit for $n = 1,\ldots,Q$. We denote the unitary matrix represented by these gates as $G^{[n, d_n]}$. Notice that the ansatz is linear in the number of qubits. Moreover, although it is sequential as shown in Fig. \ref{fig:circuit_ansatz}, it can potentially be parallelized for appropriate values of $d_n$. For example, if $d_n = 1$ for all $n$, then the ansatz is simply one layer of 1-qubit gates. Lastly, we point out that although the circuit in Fig. \ref{fig:circuit_ansatz} seems to suggest some restriction on $d_n$ (for example, $\Gate{Q-1}$ cannot have $d_n > 2$), this is, as we shall show later, not a limitation.

Armed with this ansatz, we may write down the unitary matrix of each gate in the initializer. For the first gate, we have
\begin{align}
    \nonumber
    \Gate{1} = \sum_{\substack{j_1'\ldots j_{d_1}' \\ j_1 b^1_2 \ldots b^1_{d_1}}} &\GateMatrix{1}{b^1_{d_1},\ldots,b^1_2,j_1}{j_{d_1}',\ldots,j_1'} \\
    &\quad\times \ketbra{b^1_{d_1},\ldots,b^1_2,j_1}{j_{d_1}',\ldots,j_1'}.
    \label{eq:gate1_def}
\end{align}
%
%
Consistent with our notation above, $j_n'$ above denotes the initial quantum number of the $n$\textsuperscript{th} qubit (typically 0). On the other hand, $j_n$ is the final quantum number of the $n$\textsuperscript{th} qubit after state preparation. This is distinct from $b^m_n$, which is an intermediate quantum number of the $n$\textsuperscript{th} qubit after operation of the $m$\textsuperscript{th} gate (see Fig. \ref{fig:circuit_ansatz}).
Likewise, for $n = 2,\ldots,Q-1$, $\Gate{n}$ has the form
\begin{equation}
    \Gate{n} = \sum_{ij}\GateMatrix{n}{i}{j}\ketbra{i}{j},
    \label{eq:gate_n_def}
\end{equation}
where
\begin{equation}
    i = b^n_{n + d_n - 1},\ldots,b^n_{n + 1},j_n\quad,
\end{equation}
and
%
\begin{align}
    \nonumber
    j = &
    \begin{cases}
    b^{n-1}_{n + d_n - 1},\ldots,b^{n-1}_n, \quad \text{for } d_{n-1} -1 = d_n \\
    j_{n + d_n -1}',\ldots, j_{n + d_{n-1} - 1}', b^{n-1}_{n + d_{n-1} - 2},\ldots, b^{n-1}_n,
    \end{cases}
    \\ &\hspace{3.6cm}\text{for } d_{n-1} - 1 < d_n.
    \label{eq:j_intermediate_def}
\end{align}
%
%
Notice that we have not in Eq. \eqref{eq:j_intermediate_def} defined $j$ for the case where $d_{n-1} -1 > d_n$. As we will show later, this particular case is unnecessary, but for now we have for $\Gate{Q}$
\begin{equation}
    \Gate{Q} = \sum_{j}\GateMatrix{Q}{j_Q}{j}\ketbra{j_Q}{j},
    \label{eq:gate_Q_def}
\end{equation}
where
\begin{equation}
    j =
    \begin{cases}
    b^{Q-1}_Q &, \text{ for } d_{Q-1} = 2 \\
    j_Q' &, \text{ for } d_{Q-1} = 1
    \end{cases}.
\end{equation}
%
%
The unitary matrix $U$ is then
\begin{align}
    \nonumber
    U = \sum_{\substack{j_1\ldots j_Q\\j_1'\ldots j_Q'\\\{b^1\}\ldots\{b^{Q-1}\}}} &\Gate{Q}\Gate{Q-1}\ldots\Gate{1} \\
    &\quad\times\ketbra{j_Q\ldots j_1}{j_Q'\ldots j_1'}.
\label{eq:U_def}
\end{align}
%
%
In the summation above, we have used the notation $\{b^m\}$ to mean the set of all $b$ indices with superscript equal to $m$. In other words, all the intermediate quantum numbers are summed over in Eq. \eqref{eq:U_def}. Also, since $\Gate{n}$, as defined in Eq. \eqref{eq:gate1_def}, \eqref{eq:gate_n_def}, and \eqref{eq:gate_Q_def} are $2^{d_n}$-dimensional matrices, but $U$ in Eq. \eqref{eq:U_def} is $2^Q$-dimensional, we have implicitly assumed in Eq. \eqref{eq:U_def} that each $\Gate{n}$ has been appropriately expanded via the correct Kronecker products in accordance to the qubits that it operates on.

For state preparation, we want $U\ket{0,\ldots,0}=\ket{\psi}$, where $\ket{\psi}$ is a right-canonical MPS as in Eq. \eqref{eq:mps_def}. More precisely, we want $U$ such that
\begin{align}
    \sum_{\{b^1\}\ldots\{b^{Q-1}\}} &\Gate{Q}\Gate{Q-1}\ldots\Gate{1} = \Atensor,
    \label{eq:G_sum}
\end{align}
when $j_n'=0$ for all $n$. Here, we are viewing $A$ as a tensor and the $\Gate{n}$ in their "native" space. Comparing Eq. \eqref{eq:G_sum} with Eq. \eqref{eq:TT_SVD}, we observe that in both cases, we have $Q$ free indices ($j_1,\ldots,j_Q$), $Q$ MPS cores and gates, as well as $Q-1$ non-trivial summation indices, which suggests that Eq. \eqref{eq:G_sum} can be accomplished through a mapping between each $n$\textsuperscript{th} MPS core and gate. For this mapping to occur, we need to ensure that the space spanned by $\{b^n\}$ is at least as large as $\dim(\alpha^n)$. This can be enforced by choosing
\begin{equation}
    d_n = 1 + \ceil*{\log_2 \dim(\alpha^n)}.
    \label{eq:dn_def}
\end{equation}
This choice of $d_n$ also ensures that $d_{n-1} - 1 \leq d_n \,\forall\, n$ so that our omission of the case $d_{n-1} - 1 > d_n$ in Eq. \eqref{eq:j_intermediate_def} is justified. To see this, we note that for any right-canonical MPS, the inequality Eq. \eqref{eq:dim_inequality} must hold. Taking $\log_2$ and adding one to both sides of the inequality, we have after rounding up
\begin{align}
    \nonumber
    &1 + \log_2 \ceil*{\dim(\alpha^{n-1})} \leq 2 + \ceil*{\log_2 \dim(\alpha^n)} \\
    &\implies d_{n-1} - 1 \leq d_n.
\end{align}
%

We remark here that the algorithm presented in \cite{Ran_2020} can, for the case of just a single layer, be considered to be a special case of our algorithm with $d_n = 2$ for all $n$. If the exact MPS representation of the target state has bond dimensions larger than 2, the algorithm in \cite{Ran_2020}, for the case of a single layer, approximates the target state to a matrix product state with uniform bond dimensions of 2. One consequence of this approximation in \cite{Ran_2020} is that it is unable to exactly prepare any arbitrary state with bond dimensions larger than 2, and to better approximate states with more entanglement, the algorithm in \cite{Ran_2020} needs to use multiple layers of 2-qubit gates. Conceptually, this is distinct from our approach which is able to exactly prepare any arbitrary state of $Q$ qubits with one layer of $Q - 1$ sequential $d_q$ qubit gates where $d_q$ is a variable that depends on the entanglement of the target state.

On the other hand, the circuit presented in \cite{Cramer2010} is equivalent to the case of $d_n = d$ for all $n$ where $d$ is some constant that is sufficiently large to allow a faithful representation of the target state with a MPS. However, compared to our approach, the circuit in \cite{Cramer2010} is inefficient in that it does not take into account variations in the bond dimensions of individual bonds. Obviously, for the case of a single layer, the algorithm in \cite{Ran_2020} approaches the circuit of \cite{Cramer2010} as $d_n$ increases uniformly for all $n$ and becomes increasingly capable of faithfully reproducing states with higher entanglement. Nevertheless, such a procedure is also inefficient for the same reason why \cite{Cramer2010} is not ideal.

Physically, the choice of $d_n$ in our Eq. \eqref{eq:dn_def} means that the number of qubits the $n$\textsuperscript{th} gate operates on is directly dependent on the entanglement between the sub-systems $\{j_1,\ldots,j_n\}$ and $\{j_{n+1},\ldots,j_Q\}$. For example, if $\dim(\alpha^n)=1$ $\forall n$, i.e. the target state is a product state, then $d_n=1$ $\forall n$ by Eq. \eqref{eq:dn_def}, and the circuit reduces to simply one layer of 1-qubit gates. Thus our circuit by construction only uses $d > 1$ gates when the entanglement of the target state calls for it.

It remains to prescribe the mapping from a MPS core $A^n$ to a gate $\Gate{n}$. We begin by observing that each MPS core can be reshaped into a matrix $A^n(\alpha^{n-1};\alpha^n,j_n)=A^n(i;j)$. In converting a multi-index like $\alpha^n,j_n$ into a single index $j$, we shall adopt a C-indexing convention, that is, indices on the right ($j_n$ in this case) are incremented first before those on the left ($\alpha^n$ in this case). Each $A^n$ core can thus be viewed as a matrix with $\dim(\alpha^{n-1})$ rows and $2\dim(\alpha^n)$ columns. Now by construction, $\Gate{n}$ has at least $\dim(\{b^{n-1}\}) \geq \dim(\alpha^{n-1})$ columns \footnote{For $\Gate{1}$, it has at least 2 columns from $j_1'$ and $\dim(\alpha^0)=1$, so it also has more columns than $\dim(\alpha^{n-1})$}. Furthermore, it has $2\dim(\{b^n\}) \geq 2\dim(\alpha^n)$ rows \footnote{For $\Gate{Q}$, it has 2 rows from $j_Q$ and $\dim(\alpha^Q)=1$, so its number of rows is equal to $2\dim(\alpha^n)$}. We can therefore map the $\dim(\alpha^{n-1})$ rows of $A^n(\alpha^{n-1};\alpha^n,j_n)$ into the first $\dim(\alpha^{n-1})$ columns of $\Gate{n}$, zero-padding any remaining rows. With this accomplished, any remaining columns of $\Gate{n}$ can be filled by requiring that $\Gate{n}$ is unitary, as is necessary for a quantum gate. To do this, we note that since the MPS is right-canonical, each of the $\dim(\alpha^{n-1})$, $2\dim(\alpha^n)$-dimensional row vector in $A^n$ are orthonormal to each other. Consequently, a trivial embedding of them in a potentially larger $2\dim(\{b^n\})$-dimensional space by zero-padding the extra dimensions will mean that they are still orthonormal in the larger space. We thus have $\dim(\alpha^{n-1})$ filled columns that are orthonormal vectors in a $\mathbb{C}^{2\dim(\{b^n\})}$ vector space, and $2\dim(\{b^n\}) - \dim(\alpha^{n-1})$ remaining columns to fill such that $\Gate{n}$ is unitary. Clearly, this can be accomplished by filling the still empty columns with the remaining $2\dim(\{b^n\}) - \dim(\alpha^{n-1})$ orthonormal vectors in $\mathbb{C}^{2\dim(\{b^n\})}$.

We conclude this section with a remark on the time complexity required to obtain the MPS cores required for our algorithm to prepare a state with $N$ basis states, equivalent in this context to loading a classical vector of length $N$ into a quantum computer. In analyzing the time complexity, we are mainly interested in the asymptotic complexity for large $N$, and we therefore (like the HHL algorithm) assume that the classical input data is sparse since dense inputs would require exponentially increasing classical memory requirements. As discussed in section~\ref{sec:matrix_product_states}, the MPS cores are obtained through a series of SVDs. It is therefore not difficult to show that to leading order, the complexity required to obtain the MPS cores is the sparse SVD of a $\sqrt{N}\times\sqrt{N}$ matrix, which has a time complexity of $\tilde{\mathcal{O}}(\tilde{s} + k^2 \kappa_k^4 \sqrt{N}/\epsilon^2)$, where $\tilde{s}$ is the number of non-zero elements in the $\sqrt{N}\times\sqrt{N}$ matrix, $k$ is the number of singular values requested, $\kappa_k$ is the ratio of the largest singular value to the $k$\textsuperscript{th} largest singular values of the $\sqrt{N}\times\sqrt{N}$ matrix, and $\epsilon$ is the error \cite{Allen_Zhu2016}. As a comparison, the time complexity of HHL is $\tilde{\mathcal{O}}(\log(N) s^2 \kappa^2/\epsilon)$, where $s$ is in this case the maximum number of non-zero entries per row of $A$, the $N\times N$ matrix representing the linear system of equations that HHL is attempting to ``solve'' \cite{Harrow_2009}. Similarly, $\kappa$ is here the condition number of $A$. Evidently, the time complexity of obtaining our MPS cores has a worse scaling in $N$ than HHL. However, we note that it is still better than the complexity of inverting a sparse matrix classically using the conjugate gradient method, which has a time complexity of $\mathcal{O}(N s \kappa \log(1/\epsilon))$ \cite{Harrow_2009}. Consequently, using our state preparation algorithm together with HHL can still potentially provide a speed-up over classical matrix inversion although an important caveat here is that the output of HHL does not give the full solution that classical matrix inversion provides. Finally, we note that the time complexity of computing the MPS cores scales quadratically with the number of singular values requested of the SVD. Physically, this implies that states with less entanglement have lower time complexity so that the classical cost of our algorithm is least for states with low entanglement.

\section{\label{sec:hardware_benchmark}Performance on NISQ computers}

Although our algorithm is capable of preparing arbitrary states, it is generally most efficient in terms of quantum gate operations when preparing states with low entanglement. This degree of entanglement for a $Q$ qubit system with density matrix $\rho$ and qubits $\{j_1,\ldots,j_Q\}$ can be quantified with the mean normalized bipartite entropy, which we define as
\begin{equation}
    \bar{S} = \left\langle\frac{S(\rho_n)}{\min(n, Q - n)} \right\rangle = \left\langle - \frac{\Tr(\rho_n \log_2 \rho_n)}{\min(n, Q-n)} \right\rangle,
    \label{eq:Sbar_def}
\end{equation}
where $\rho_n \equiv \Tr_n (\rho)$ is the reduced density matrix of the sub-partition $\{j_1,\ldots,j_n\}$, and the average is over all sub-partitions $\{j_1\}$, $\{j_1, j_2\}$, $\ldots$, $\{j_1,\ldots,j_{Q-1}\}$. Note that $S(\rho_n)/\min(n, Q - n)$ is just the normalized bipartite von Neumann entropy, which is a measure of the entanglement between the bipartitions $\{j_1,\ldots,j_n\}$ and $\{j_{n+1},\ldots,j_Q\}$, with zero indicating a separable state and one indicating a maximally entangled state. $\bar{S}$, which ranges between zero and one, is therefore the average normalized bipartite von Neumann entropy over all bipartitions of the system. Generally speaking, our algorithm yields the most savings in terms of gate count and circuit depth compared to other deterministic methods when $\bar{S}$ is small.

\begin{figure*}[!ht]
    \centering
    \includegraphics[width=0.98\textwidth]{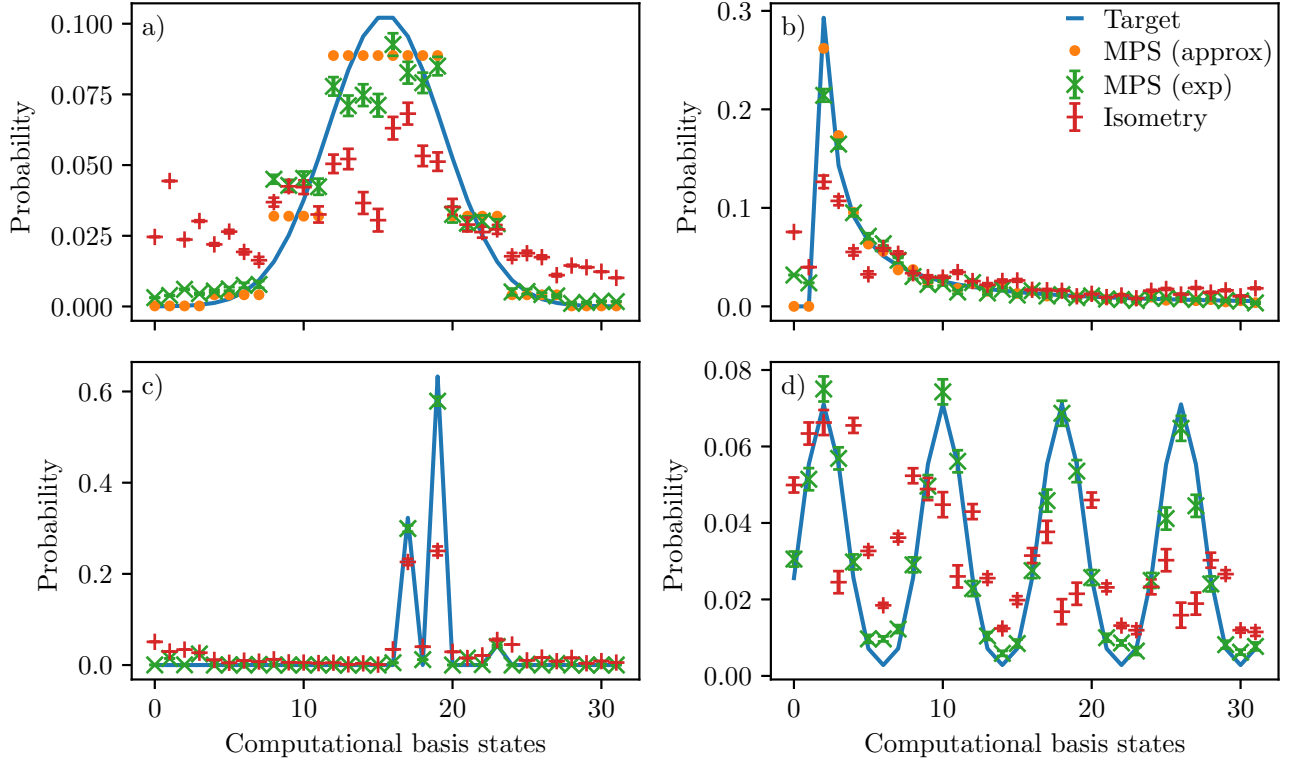}
    \caption{(color online) Preparation of various probability distributions on IBM's quantum computers: (a) normal distribution, (b) log-normal distribution, (c) a "random" distribution (see text for more details), and (d) a sinusoidal distribution. Blue solid lines demarcate the target probability distribution, while orange dots (if present) show the locally optimal approximate state with lower entanglement. Green crosses show experimental results obtained with our routine while red pluses show those obtained using isometric decomposition. Error bars give the theoretical 1-sigma spread of the measured probabilities.}
    \label{fig:dist_plot}
\end{figure*}

An example of a target state with a relatively low mean normalized bipartite entropy is the sinusoidal probability distribution shown in Fig. \ref{fig:dist_plot}d, with $\bar{S} = 0.076$. In this case, we were able to, without any approximation, prepare the target distribution with an order of magnitude less CNOT gates and with a circuit that is shallower by about three times compared to isometric decomposition \cite{Iten_2016} (details in Table \ref{tab:benchmark}). Not surprisingly, our algorithm was also able to achieve a higher fidelity, which can be visualized in Fig. \ref{fig:dist_plot}d that shows the target distribution in a solid blue line, as well as the measured probability distribution obtained via our algorithm (green crosses) and isometric decomposition (red pluses). Error bars shown in this and other plots give the expected 1-sigma deviation of the measured probabilities based on the number of shots and theoretical probability of obtaining each state. In all our tests, we simply chose the least busy quantum hardware at execution time without special consideration for the error rate and qubit connectivity and we used the standard transpilation routine on Qiskit \cite{Qiskit} with the highest level of optimization. Although the results summarized in Table \ref{tab:benchmark} were all obtained from quantum devices with a one-dimensional qubit layout, we have also obtained similar results using backends like \textit{ibmq\_lima} that have two-dimensional layouts.

For states with larger entanglement, optimally approximate states can be found using the procedure outlined in appendix~\ref{sec:matrix_product_states}, and using this approximate state, our method can sometimes still achieve better fidelities than state preparation via existing deterministic methods. In the first row of Fig. \ref{fig:dist_plot}, we compare our preparation of a normal (a) and log-normal (b) distribution, which are both of interest to the quantum finance community \cite{Rebentrost_2018, Woerner_2019, Egger_2020, Stamatopoulos_2020}. In both cases, the target probability distributions are plotted as blue solid lines while their matrix product state approximations are given by orange dots. As discussed in appendix~\ref{sec:matrix_product_states}, these are states that are optimally (in the Frobenius norm) approximate to the target state but with lower entanglement. To ensure a fair comparison, the isometric state preparation shown in Fig. \ref{fig:dist_plot}a and Fig. \ref{fig:dist_plot}b was also used to prepare the same approximate MPS target state as our algorithm. However, as Table \ref{tab:benchmark} shows, there is actually little difference in the experimental fidelity of the isometric method when it prepares the exact target state instead. This is not surprising since isometric decomposition does not specifically optimize based on the entanglement of the target state. We emphasize that although our algorithm prepared an approximate state, it nevertheless performs substantially better than an exact state preparation using isometric decomposition as the results in Table \ref{tab:benchmark} shows. In particular, for the case of the normal distribution in Fig. \ref{fig:dist_plot}a, our method used more than an order of magnitude less CNOT gates, with about 2.5 times shallower depth, and achieved a significantly higher fidelity to the target state.

\begin{table*}[!ht]
    \centering
    \begin{tabular*}{0.99\textwidth}{@{\extracolsep{\fill}} cccccccccc}
        \hline\hline
        Distribution & Hardware & Method & CNOT & $R_z$ & SX & Circuit & Fidelity & Fidelity & Mean normalized  \\
         & & & & & & depth & (exp) & (theory) & bipartite entropy \\
        \hline
        Normal & ibmq\_santiago & Ours (approx) & 5 & 35 & 30 & 44 & 0.980 & 0.990 & 0.283 \\
         & & Isometry (approx) & 71 & 62 & 55 & 122 & 0.872 & 0.990 & 0.283 \\
         & & Isometry (exact) & 77 & 65 & 49 & 123 & 0.864 & 1 & 0.322 \\
        \hline
        Log-normal & ibmq\_santiago & Ours (approx) & 9 & 45 & 44 & 57 & 0.967 & 0.998 & 0.192 \\
         & & Isometry (approx) & 73 & 78 & 68 & 144 & 0.907 & 0.998 & 0.192 \\
         & & Isometry (exact) & 63 & 66 & 51 & 129 & 0.917 & 1 & 0.206 \\
         \hline
        Random & ibmq\_santiago & Ours (exact) & 2 & 17 & 17 & 22 & 0.959 & 1 & 0.014 \\
         & & Isometry (exact) & 70 & 72 & 61 & 140 & 0.717 & 1 & 0.014 \\
        \hline
        Sinusoidal & ibmq\_manila & Ours (exact) & 5 & 30 & 28 & 40 & 0.995 & 1 & 0.076 \\
         & & Isometry (exact) & 57 & 65 & 61 & 118 & 0.921 & 1 & 0.076 \\
        \hline\hline
    \end{tabular*}
    \caption{Benchmark of our state preparation algorithm compared to isometric decomposition. Gate (CNOT, $R_z$, and SX) counts and circuit depth for each method, as well as their experimentally measured fidelity to the target state are tabulated. When preparing an approximate state is used, the fidelity of the approximate state to the exact state is listed under "Fidelity (theory)" (an entry with 1 means the exact state was used). The mean normalized bipartite entropy of the exact target state, defined in Eq. \eqref{eq:Sbar_def}, is also given.}
    \label{tab:benchmark}
\end{table*}

As the number of qubits grows, any classical input data to a quantum algorithm like HHL must also be sparse since dense classical inputs would require exponentially increasing memory requirements that does not exist. It is therefore typically assumed that classical inputs to quantum algorithms like HHL are sparse, and it is hence interesting to see if our method can generate substantial savings in such cases. To test this, we randomly generated a vector with a sparsity of 0.9 where both the location and magnitude of the non-zero elements were randomly chosen. Although not every sparse vector will necessarily have low $\bar{S}$, some sparse vectors may have very low entropy, and in this case, our algorithm can enable significantly better results over isometric state preparation. For example, the sparse vector state visualized in Fig. \ref{fig:dist_plot}c has a very low $\bar{S}$ of 0.014, and in this case, we were able to exactly prepare the state with just 2 CNOT gates, compared to 70 with an isometric decomposition. This is a striking demonstration of the utility of entanglement-sensitive routines like ours. With a circuit depth that is more than six times shallower than isometric decomposition, we were able to achieve a significantly higher fidelity of 0.959 over 0.717.

\begin{figure*}[!ht]
    \centering
    \includegraphics[width=0.98\textwidth]{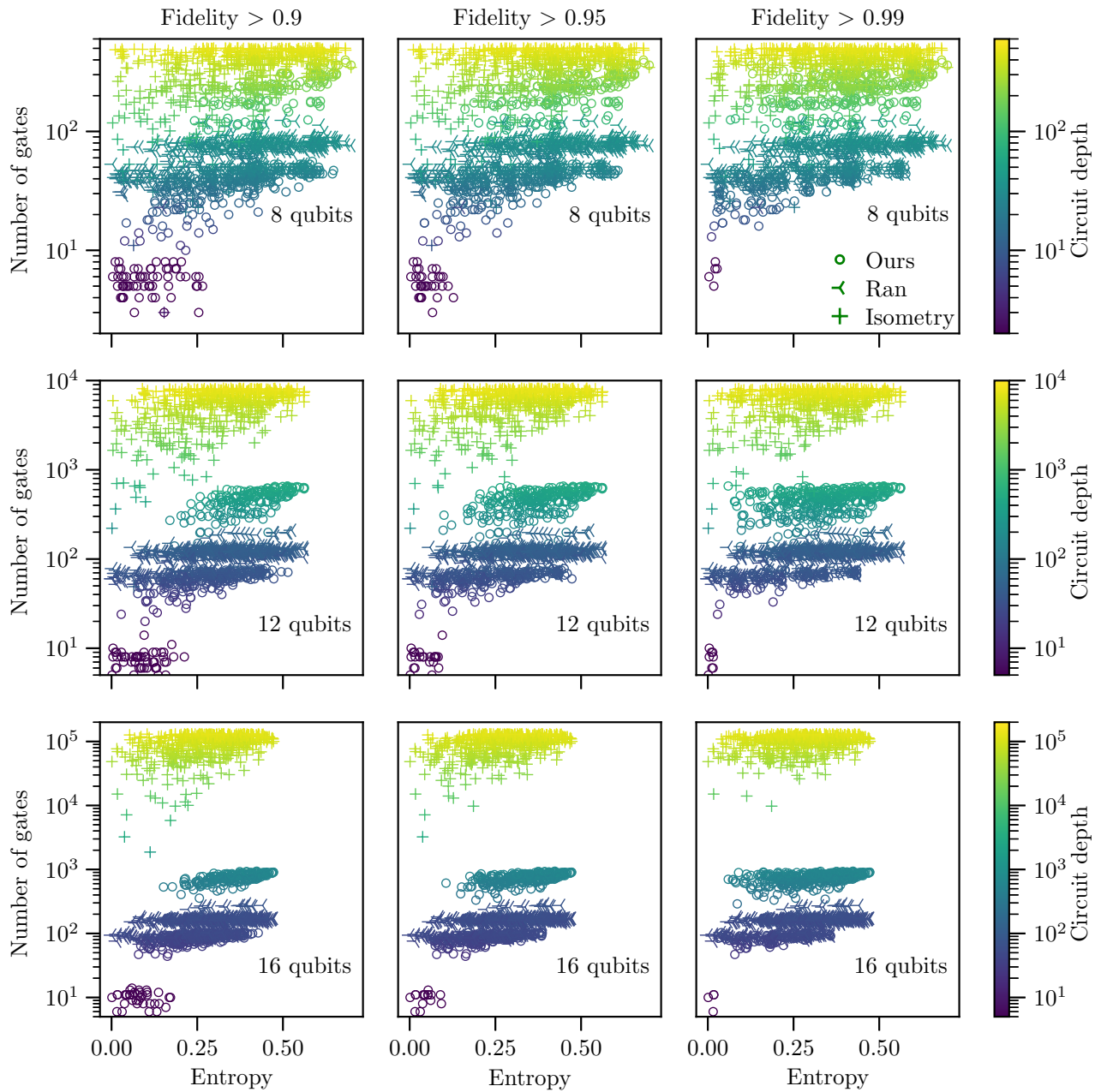}
    \caption{(color online) Minimum number of basis gates (vertical axis) and circuit depth (color of markers) when preparing states with a variety of mean normalized bipartite entropies (as defined in Eq. \eqref{eq:Sbar_def}) so as to achieve theoretical fidelities $>$ 0.9, 0.95, and 0.99 using our method, Ran's method \cite{Ran_2020}, and isometric decomposition \cite{Iten_2016}. As in Fig. \ref{fig:dist_plot}, the isometric decomposition prepares the same approximate MPS states rather than the exact states for a fair comparison.}
    \label{fig:sparse_compare}
\end{figure*}

As we demonstrated on NISQ computers in Fig.~\ref{fig:dist_plot} and Table~\ref{tab:benchmark}, having the ability to prepare optimally approximate states is important in the NISQ era since one may obtain higher fidelities to the target state by preparing approximate states instead of the exact state. In our algorithm, this is because such approximate states have lower entanglement and require fewer entangling gates to prepare, which results in reduced experimental decoherence on NISQ computers. One feature of our algorithm is its ability to successively approximate the target state so that successive approximate states can, at the expense of their fidelity to the target state, be prepared with fewer gates. From a practical point of view, one can therefore run our algorithm till the (theoretical) fidelity to the target state falls below some acceptable threshold, and then choose the circuit produced by the preceding iteration as the one to use. We emphasize that while this theoretical fidelity is an upper limit to the achievable fidelity, a circuit produced by our algorithm may on NISQ computers nevertheless experimentally outperform another algorithm that gives a higher theoretical fidelity but suffers from significantly more decoherence experimentally due to its deeper circuit and use of considerably more gates.

\section{\label{sec:simulation_benchmark}Performance in numerical simulations}

In Fig.~\ref{fig:sparse_compare}, we plot the minimum number of basis gates and circuit depth (color of markers) required to prepare a randomly generated state with fidelities above 0.9, 0.95, and 0.99, using our algorithm (circular markers), Ran's algorithm \cite{Ran_2020}, and isometric decomposition \cite{Iten_2016} (`+' markers). The entropies ($\bar{S}$ from Eq. \eqref{eq:Sbar_def}) of these states is plotted on the horizontal axis. To test the scaling of the algorithms, we performed the study for 8, 12, and 16 qubits. Clearly, isometric decomposition consistently performs worse in requiring more gates and deeper circuits as compared to both Ran's algorithm and our method. To ensure a fair comparison, the isometric decomposition method here was used to prepare the same approximate state as that produced by our algorithm. We checked that preparing the exact state will generally require more resources even for isometric decomposition, and we have therefore for clarity not shown this poorer result in Fig.~\ref{fig:sparse_compare}.



It is interesting to contrast Ran's circuit, which consists of successive layers of sequential 2-qubit gates, with our algorithm. As Fig~\ref{fig:sparse_compare} demonstrates, Ran's method generally requires a consistent number of gates and depth with little dependence on the entropy of the target state. On the other hand, our approach shows markedly better performance at low entropies, and similar to worse performance at higher entropies, with increasing frequency of better performance at decreasing fidelity requirements. We highlight that these are theoretical fidelity requirements, and that on NISQ computers, our algorithm that prepares a state with theoretical fidelity $> 0.9$ can experimentally outperform another algorithm with theoretical fidelity $> 0.99$ by using significantly fewer gates and by having a much shallower circuit (see, for example, the tests in section~\ref{sec:hardware_benchmark} and the results in Table~\ref{tab:benchmark}). 

Physically, the difference between our algorithm and Ran's method is that whereas Ran uses successive layers of sequential 2-qubit gates to increasingly approximate the target state, we use a single layer of sequential $d_q$-qubit gates, where $d_q$ is an entanglement-dependent variable and $q = 1,\ldots,Q-1$ for a $Q$ qubit system. Since each gate operates on a variable number of qubits, and $d_q$ is sensitive to the entanglement of the state, we can prepare low entanglement states with very few gates. On the hand, because Ran's method can only use discrete layers of 2-qubit gates (this can be observed in Fig.~\ref{fig:sparse_compare} as discrete bands in the of number of gates for the Ran method), it is unable to prepare low entanglement states as efficiently as our algorithm. Finally, we note that our method can, in principle, prepare any state exactly, whereas Ran's method can only asymptotically become exact with increasing number of layers. Nevertheless, we acknowledge that this ability to prepare states exactly is largely moot in the NISQ era due to environmental decoherence.

\section{Conclusion\label{sec:conclusion}}

Arbitrary quantum states are difficult to prepare with high fidelities on today's NISQ computers. Due to noise from multi-qubit entangling gates, deep quantum circuits with a large number of such gates rapidly become incoherent and useless. Although analytical methods to exactly prepare arbitrary states have been known for a while, these approaches suffer from deep circuits and high entangling gate counts that increase exponentially with the number of qubits. This translates in practice to significant decoherence of the qubits that makes them unusable for further computation. In response to this, variational circuits with parametrized gates have since been proposed for state preparation. Nevertheless, these circuits are not a panacea and are instead plagued with their own problems. For example, finding the loss function's global minimum in these methods is highly challenging due to the curse of dimensionality, gradients that vanish exponentially with the number of qubits, and a proliferation of local minimas in the loss function's landscape. More importantly, there is no obvious guiding principle for the selection of the circuit's topology, and there is no guarantee that any given parametrized circuit will actually be able to prepare the state exactly. Furthermore, none of these existing approaches take into account the fact that states with less entanglement should in theory be easier to prepare with less entangling gates.

In this work, we use a state preparation circuit that uses only the same number of nearest-neighbor gates as the number of qubits, which minimizes the number of expensive SWAP operations due to limited qubit connectivity. Although each of these nearest-neighbor gates may be an expensive multi-qubit gate that further decomposes into multiple one and two-qubit gates, our routine only uses such gates sparingly when the entanglement of the target state calls for them. In cases where a target state (or a sub-system of it) is separable, our circuit automatically takes advantage of this by utilizing single qubit gates and parallelizing gate operations where possible.

In short, our state preparation routine allows states with less entanglement to be prepared more efficiently compared to other deterministic but entanglement-insensitive methods such as isometric decomposition. For states with low entanglement, our tests on actual quantum computers hosted on IBM's cloud service show that our circuit is capable of achieving higher state fidelities compared to standard isometric decomposition by utilizing an order of magnitude less CNOT gates and having a significantly shallower circuit. For states with higher entanglement, we can still sometimes achieve significantly better performance by preparing an optimally approximate state with lower entanglement, including states that represent a normal and log-normal distribution. Finally, we emphasize that unlike variational approaches, our method does not require challenging minimizations in a high dimensional space where the existence of a solution is uncertain, and exponentially vanishing gradients and a multitude of local minima make progress towards a global minimum difficult. Our state preparation algorithm is therefore a valuable tool in the NISQ era for preparing arbitrary states with acceptable fidelities without having to incur significant classical optimization costs.

\begin{acknowledgments}
The authors acknowledge Aravind Anthur and Leonid Krivitsky's contributions in securing grant QEP-SF2 from the Quantum Engineering Programme in Singapore that provided financial support for this work. Numerical simulations were conducted with the resources of A*STAR Computational Resource Center (A*CRC), with the Python library Dask \cite{dask} used for parallelization. Quantum circuit simulations and executions were performed using Qiskit \cite{Qiskit}.
\end{acknowledgments}


 
\providecommand{\noopsort}[1]{}\providecommand{\singleletter}[1]{#1}%

\end{document}